\documentclass[aps,prd,reprint,nofootinbib,superscriptaddress]{revtex4-2}

\usepackage[T1]{fontenc}
\usepackage[utf8]{inputenc}
\usepackage{lmodern}
\usepackage{amsmath,amssymb,amsfonts,bm,mathtools}
\usepackage{booktabs}
\usepackage{hyperref}
\usepackage{xcolor}

\hypersetup{
  colorlinks=true,
  linkcolor=blue,
  citecolor=blue,
  urlcolor=blue
}

\newcommand{\ve}[1]{\bm{#1}}
\newcommand{\diag}{\mathrm{diag}}
\newcommand{\EC}{E_{\rm C}}
\newcommand{\PC}{P_{\rm C}}

\newcommand{\Lcal}{\mathcal L}

\newcommand{\Eone}{E_0^{(1)}}

\begin{document}

\title{Effective-metric formulation of Casimir energies in nonlinear scalar and electromagnetic theories}

\author{C. A. Escobar}
\email{carlos.escobar@xanum.uam.mx}
\affiliation{Departamento de F\'isica, Universidad Aut\'onoma Metropolitana-Iztapalapa, San Rafael Atlixco 186, Ciudad de M\'exico 09340, M\'exico}

\begin{abstract}
We study the Casimir effect in nonlinear field theories through the effective geometries that \mbox{govern} their linearized fluctuations. Previous analyses of Lorentz-violating scalar fields showed that a constant kinetic background modifies the parallel-plate Casimir energy by a rescaling of the plate separation and an overall determinant factor. We show that this structure is not merely a consequence of diagonalizing the reduced Green function. It follows from a common Schur-complement structure: after Fourier reduction parallel to the plates, the same reduced quadratic form controls the spectral denominator of the reduced Green function and the numerator generated by the energy-density insertion. This observation allows the Lorentz-violating scalar result to be used as an effective-metric prescription for regular fluctuation sectors arising from the linearization of nonlinear theories around constant backgrounds. In nonlinear scalar theories, the effective tensor is the Hessian of the Lagrangian evaluated on a constant-gradient background. In nonlinear electrodynamics $\mathcal{L}(\mathcal{F})$, a constant magnetic background splits the fluctuations into an ordinary Maxwell branch and an extraordinary optical branch. For this electromagnetic sector, we compute the parallel-plate Casimir energy both by direct mode summation and by applying the effective-metric formula branch by branch, finding exact agreement. The resulting energy depends on the orientation of the magnetic background relative to the plates, providing a concrete anisotropic Casimir response in a regular nonlinear electromagnetic sector.
\end{abstract}

\maketitle

\section{Introduction}
\label{sec:introduction}

The Casimir effect is one of the most direct manifestations of vacuum fluctuations in quantum field theory. In the parallel-plate geometry, the boundary conditions restrict the spectrum of quantum fluctuations and generate a finite renormalized vacuum energy after subtracting the corresponding unbounded contribution \cite{Casimir1948,Plunien1986,Milton2001,Bordag2009}. For a single massless scalar field with Dirichlet boundary conditions, the energy per unit area is
\begin{equation}
\label{eq:single_branch_energy_intro}
\Eone(L)=-\frac{\pi^2}{1440L^3}.
\end{equation}
The standard electromagnetic result for two perfectly conducting plates is twice this value, reflecting the two physical photon polarizations in the Lorentz-invariant theory \cite{BrownMaclay1969,Milton2001,Bordag2009}.

The Casimir effect is also a useful probe of modified fluctuation operators. If the kinetic operator is not Lorentz invariant, the vacuum energy may depend not only on the plate separation, but also on the orientation of the background structures relative to the plates. This issue has been explored in several Lorentz-violating scalar and electromagnetic settings \cite{Cruz2017,Cruz2018,MartinRuiz2017LocalEM,Escobar2020Local,Escobar2020PLB,MartinRuiz2020Sphere,EscobarRuiz2021Cylinder,DantasMotaBezerra2023HigherDerivatives,BezerraCruz2023ConstantVectors,CruzBezerraPetrov2019Fermionic}. A particularly useful nonperturbative result was obtained for a real scalar field governed by the quadratic Lagrangian
\begin{equation}
\label{eq:h_lagrangian_intro}
\Lcal_h=\frac12\left(h^{\mu\nu}\partial_\mu\phi\,\partial_\nu\phi-m^2\phi^2\right),
\end{equation}
where $h^{\mu\nu}$ is a constant, symmetric, nondegenerate tensor of Lorentzian signature. For two parallel plates normal to a spatial direction $n$, the global Casimir energy and pressure take the form
\begin{align}
\label{eq:plb_formula_intro}
\EC^{(h)}(L)
&=
\sqrt{\frac{h^{nn}}{\det h}}\,
E_0\left(\frac{L}{\sqrt{-h^{nn}}}\right),
\\
\label{eq:plb_pressure_intro}
\PC^{(h)}(L)
&=
\frac{1}{\sqrt{-\det h}}\,
P_0\left(\frac{L}{\sqrt{-h^{nn}}}\right),
\end{align}
in the regular sector where $h^{nn}<0$. Here $E_0$ and $P_0$ denote the corresponding Lorentz-invariant scalar expressions.

The derivation of Eqs.~\eqref{eq:plb_formula_intro} and \eqref{eq:plb_pressure_intro} in Ref.~\cite{Escobar2020PLB} proceeds through the Green-function method. One first exploits translational invariance along the directions parallel to the plates and Fourier transforms only in those coordinates. The problem is then reduced to a one-dimensional boundary-value problem in the normal coordinate. The reduced Green function contains a phase associated with the normal-parallel mixing of the tensor $h^{\mu\nu}$ and a quadratic form in the frequency and transverse momenta. The vacuum expectation value of the stress tensor is evaluated by point splitting, and the physical energy and pressure are obtained after the standard subtraction of the free or exterior contributions. A suitable linear transformation of the frequency-momentum variables, followed by a Wick rotation, brings the relevant Green-function integrals to the Lorentz-invariant form, at the price of a determinant factor and a rescaling of the plate separation.

This procedure gives closed expressions for the energy and pressure, but it also raises a question that becomes important once the tensor \(h^{\mu\nu}\) is interpreted as an effective background. The diagonalization of the quadratic form controlling the spectral part of the reduced Green function explains the appearance of an effective separation, but it does not by itself explain why the stress-tensor numerator transforms compatibly. The first part of this work addresses this point in a general quadratic setting and shows that the relevant numerator is fixed by the same Schur complement that governs
the spectral part of the reduced Green function.

This observation allows one to reinterpret Eqs.~\eqref{eq:plb_formula_intro} and \eqref{eq:plb_pressure_intro} as effective-metric formulas. The tensor $h^{\mu\nu}$ need not be introduced as an external Lorentz-violating background from the outset. It may arise as the kinetic tensor governing linearized fluctuations around a nontrivial classical background. This is immediate in nonlinear scalar theories: for a Lagrangian depending on the standard kinetic invariant, the Hessian with respect to field gradients plays the role of the effective metric when evaluated on a constant-gradient background. In this way, the Casimir response probes the geometry of the fluctuation operator rather than only the microscopic form of the original nonlinear theory.

Nonlinear electrodynamics provides a more restrictive and more informative test of this idea. Nonlinear electromagnetic theories have long been studied as effective descriptions of quantum-vacuum polarization, Born-Infeld-type dynamics, and modified light propagation in external fields \cite{BornInfeld1934,HeisenbergEuler1936,Schwinger1951,BialynickaBirula1970,Boillat1966,Boillat1970,Novello2000,Novello2003,ObukhovRubilar2002,DeMelo2014,RussoTownsend2023}. In this case the quadratic expansion does not generically produce a single second-rank metric. It first produces an effective constitutive tensor, and the relevant optical metrics appear only after solving the polarization problem. We focus on the regular Lagrangian class
\begin{equation}
\label{eq:lf_intro}
\mathcal{L}=\mathcal{L}(\mathcal{F}),\qquad
\mathcal{F}=\frac14F_{\mu\nu}F^{\mu\nu}
=\frac12(\ve B^2-\ve E^2),
\end{equation}
with metric convention $\eta_{\mu\nu}=\diag(+,-,-,-)$ and Maxwell theory given by $\mathcal{L}=-\mathcal{F}$. Around a constant magnetic background, the physical electromagnetic spectrum splits into an ordinary Maxwell branch and an extraordinary anisotropic branch. The latter is described by an optical metric whose spatial eigenvalues depend on the nonlinear response of the theory evaluated on the background.

The electromagnetic sector will be used as a direct consistency check of the effective-metric prescription. We compute the parallel-plate Casimir energy in two independent ways. First, we quantize the physical ordinary and extraordinary branches and perform the corresponding mode summation. Second, we identify the optical metrics associated with the two branches and apply Eq.~\eqref{eq:plb_formula_intro} to each regular branch separately. For magnetic backgrounds either normal or parallel to the plates, both procedures agree exactly. The resulting energy depends on the relative orientation between the magnetic background and the plates, giving an explicit realization of an anisotropic Casimir response in a regular nonlinear electromagnetic sector.

The analysis is restricted throughout to regular, nondegenerate optical sectors. We assume that the background is constant, that the linearized coefficients are constant, and that the branch decomposition is compatible with the perfect-conductor boundary conditions in the two orientations considered. Cases in which the constitutive tensor loses rank, an optical metric becomes degenerate, or the number of propagating degrees of freedom changes require a separate constrained treatment. Such singular sectors arise naturally in some Pleba\'nski formulations of nonlinear electrodynamics with Lorentz-breaking magnetic vacua, but they are not part of the present $\mathcal{L}(\mathcal{F})$ construction \cite{EscobarPotting2020,Escobar2026MagneticCasimir,PlacidoFlores2026MagneticVacua}.

The paper is organized as follows. Section~\ref{sec:general_quadratic} develops the general quadratic reduction in a parallel-plate geometry. Section~\ref{sec:energy_insertion} derives the energy insertion and the reduced Ward identity. Section~\ref{sec:effective_metric_formula} obtains the effective-metric Casimir formula and recovers the Lorentz-violating scalar result as a special case. Section~\ref{sec:scalar} applies the construction to nonlinear scalar theories. Section~\ref{sec:nled_linearization} derives the optical branches of \(\mathcal{L}(\mathcal{F})\) electrodynamics around a constant magnetic background. Sections~\ref{sec:mode_sum} and \ref{sec:branchwise} compare the direct mode-sum calculation with the branchwise effective-metric computation. Section~\ref{sec:example} presents a weakly nonlinear example, and Sec.~\ref{sec:conclusions} summarizes the results and limitations.

\section{General quadratic reduction in a parallel-plate geometry}
\label{sec:general_quadratic}

The result obtained for a Lorentz-violating scalar field with a prescribed tensor \(h^{\mu\nu}\) is a consequence of a more general property of quadratic fluctuation operators. We therefore begin with a generic regular fluctuation sector described by the quadratic action
\begin{equation}
\label{eq:general_quadratic_action}
S_2[\psi]
=
\frac12
\int d^4x\,
\left(
H^{\mu\nu}\partial_\mu\psi\,\partial_\nu\psi
-
m^2\psi^2
\right),
\end{equation}
where \(H^{\mu\nu}\) is constant, symmetric, nondegenerate, and has Lorentzian signature. The tensor \(H^{\mu\nu}\) may represent an externally prescribed Lorentz-violating background, as in previous scalar analyses, or it may arise as the effective kinetic tensor of fluctuations around a nontrivial background. The derivation below depends only on the quadratic structure of the fluctuation operator and not on the microscopic origin of \(H^{\mu\nu}\).

We consider two parallel plates located at \(z=0\) and \(z=L\). The
spatial normal direction is denoted by \(z\), while the coordinates
tangent to the plate worldvolume are labeled by indices \(A,B=0,1,2\).
Thus \(x^\mu=(x^A,z)\), with \(x^A=(t,x,y)\). The spatial directions
along the plates are \(x\) and \(y\), whereas the time coordinate is
included in \(x^A\) because the static plate configuration preserves
time-translation invariance. We assume that the normal direction is
spacelike with respect to the effective kinetic tensor,
\begin{equation}
\label{eq:normal_spacelike_condition}
H^{33}<0 .
\end{equation}

Following Ref.~\cite{Escobar2020PLB}, we define the time-ordered
two-point function of the fluctuation field as
\begin{equation}
\label{eq:green_definition_general}
G(x,x')
=
-i\langle 0|T\psi(x)\psi(x')|0\rangle .
\end{equation}
The corresponding Green-function equation is
\begin{equation}
\label{eq:general_green_equation}
\left(
H^{\mu\nu}\partial_\mu\partial_\nu+m^2
\right)
G(x,x')
=
\delta^{(4)}(x-x') .
\end{equation}

The plates break translational invariance only in the normal direction.
We therefore Fourier transform along \(t,x,y\) only, using
\begin{equation}
\label{eq:kappa_convention}
e^{i\kappa_A(x^A-x'^{A})}
=
e^{-i\omega(t-t')+i\mathbf{k}_\perp\cdot(\mathbf{x}_\perp-\mathbf{x}'_\perp)} ,
\end{equation}
so that
\begin{equation}
\label{eq:kappa_definition}
\kappa_A=(-\omega,k_x,k_y),
\qquad
x^A=(t,x,y).
\end{equation}
With this convention, derivatives along the directions parallel to the plates act as
\begin{equation}
\label{eq:parallel_derivative_rule}
\partial_A\longrightarrow i\kappa_A .
\end{equation}
The Green function is then written as
\begin{equation}
\label{eq:general_green_fourier}
\begin{aligned}
G(x,x')
&=
\int\frac{d\omega}{2\pi}
\int\frac{d^2\mathbf{k}_\perp}{(2\pi)^2}\,
e^{-i\omega(t-t')}
e^{i\mathbf{k}_\perp\cdot(\mathbf{x}_\perp-\mathbf{x}'_\perp)}
\\
&\hspace{1.2cm}\times
g(z,z';\omega,\mathbf{k}_\perp).
\end{aligned}
\end{equation}

Substitution into Eq.~\eqref{eq:general_green_equation} gives a one-dimensional problem in the normal coordinate for the reduced Green function $g(z,z';\omega,\mathbf{k}_\perp)$. Let us define the reduced normal operator
\begin{equation}
\label{eq:reduced_operator_def}
\mathcal{D}_z(\kappa)
=
H^{33}\partial_z^2
+
2iH^{3A}\kappa_A\partial_z
-
H^{AB}\kappa_A\kappa_B
+
m^2.
\end{equation}
Then the reduced Green equation satisfies
\begin{equation}
\label{eq:reduced_operator_general}
\mathcal{D}_z(\kappa)\,
g(z,z';\kappa)
=
\delta(z-z') .
\end{equation}

This equation is the starting point of the reduction. The term proportional to \(\partial_z\) appears only when the tensor \(H^{\mu\nu}\) mixes the normal direction with the directions parallel to the plates, namely when \(H^{3A}\neq0\). It is not a boundary interaction; it is a bulk effect of using a kinetic tensor that is not block diagonal with respect to the decomposition \((x^A,z)\).

To identify the normal wave numbers of the homogeneous solutions, one may
locally set \(g\sim e^{iqz}\). Multiplying the homogeneous equation by an
irrelevant overall sign, the associated quadratic polynomial is
\begin{equation}
\label{eq:general_symbol}
\mathcal{P}(q;\kappa)
=
H^{33}q^2
+
2H^{3A}\kappa_A q
+
H^{AB}\kappa_A\kappa_B
-
m^2 .
\end{equation}
The polynomial \(\mathcal{P}(q;\kappa)\) is not an additional dynamical object. It is the algebraic symbol of the reduced bulk operator for fixed parallel momentum \(\kappa_A\). Its roots determine the possible normal wave numbers before imposing the boundary conditions. Once the plates are imposed, these homogeneous solutions are combined to satisfy the boundary conditions at \(z=0\) and \(z=L\). Thus \(\mathcal{P}(q;\kappa)\) controls the part of the reduced Green function that determines the
normal mode spectrum between the plates.

The mixed term \(2H^{3A}\kappa_A q\) can be reorganized by completing the square:
\begin{equation}
\label{eq:general_square_completion}
\mathcal{P}(q;\kappa)
=
H^{33}
\left(
q+
\frac{H^{3A}\kappa_A}{H^{33}}
\right)^2
+
\kappa_A M^{AB}\kappa_B
-
m^2 ,
\end{equation}
where
\begin{equation}
\label{eq:schur_complement}
M^{AB}
=
H^{AB}
-
\frac{H^{A3}H^{B3}}{H^{33}} .
\end{equation}
The tensor \(M^{AB}\) is the Schur complement of the normal block \(H^{33}\) in \(H^{\mu\nu}\). It is the effective quadratic tensor that remains in the parallel variables once the normal-parallel mixing has been accounted for. We denote the remaining reduced bulk symbol by
\begin{equation}
\label{eq:Pi_red_definition}
\Pi_{\rm red}(\kappa)
=
\kappa_A M^{AB}\kappa_B-m^2 .
\end{equation}
This object will be central in what follows: it is the quadratic form that governs the dependence of the reduced Green function on \(\omega,k_x,k_y\) after the normal momentum has been shifted.

The same reduction can be implemented directly at the level of the
differential equation. It is useful to introduce
\begin{equation}
\label{eq:general_xi0}
\xi_0(\kappa)
=
-\frac{H^{3A}\kappa_A}{H^{33}} ,
\end{equation}
which measures the displacement of the normal momentum induced by the
mixed components \(H^{3A}\). The completion of the square then shows
that the original normal momentum is replaced by \(q-\xi_0\). In coordinate space this shift is implemented by extracting a phase from the reduced Green function:
\begin{equation}
\label{eq:general_phase_factor}
g(z,z';\kappa)
=
e^{-i\xi_0(z'-z)}
\bar g(z,z';\kappa).
\end{equation}
The equivalence is seen by differentiating this expression. Since the phase depends on \(z\),
\begin{equation}
\label{eq:derivative_shift_phase}
\partial_z g
=
e^{-i\xi_0(z'-z)}
\left(
\partial_z+i\xi_0
\right)
\bar g .
\end{equation}
The phase factor therefore realizes, at the level of the differential equation, the same normal-momentum shift obtained by completing the square. The normal-parallel mixing is not lost in this reduction; it is encoded in the phase of the reduced Green function.

After this phase is extracted, the function \(\bar g\) satisfies a second-order equation without a first derivative in \(z\). The normal dependence of its homogeneous solutions is controlled by the shifted normal wave number. If \(\bar g\sim e^{i\tilde q z}\), then the full Green function carries the original normal momentum \(q=\tilde q+\xi_0\), or equivalently \(\tilde q=q-\xi_0\).

The normal dependence of the reduced Green function, and hence the
normal wave number that enters the boundary-value problem, is therefore
governed by
\begin{equation}
\label{eq:general_xi1}
\xi_1^2(\kappa)
=
-\frac{1}{H^{33}}
\left(
\kappa_A M^{AB}\kappa_B-m^2
\right)
=
-\frac{\Pi_{\rm red}(\kappa)}{H^{33}} .
\end{equation}
Thus \(\xi_1\) is the shifted normal wave number that remains after the
normal-parallel mixing has been absorbed into the phase. It is this
quantity, rather than the original \(q\), that enters the one-dimensional
boundary Green function.

For Dirichlet boundary conditions at \(z=0\) and \(z=L\), the reduced Green function can then be written as
\begin{equation}
\label{eq:general_reduced_green}
g(z,z';\kappa)
=
e^{-i\xi_0(z'-z)}
\frac{
\sin(\xi_1 z_<)\,
\sin[\xi_1(z_>-L)]
}{
H^{33}\xi_1\sin(\xi_1 L)
},
\end{equation}
where \(z_>\) and \(z_<\) denote the greater and lesser of \(z\) and \(z'\). The sine functions are the usual standing-wave factors of the one-dimensional Green function between two Dirichlet plates. The phase factor contains the normal-parallel mixing, while \(\xi_1\) and the denominator are controlled by the Schur complement \(M^{AB}\). In the Lorentz-invariant limit \(H^{\mu\nu}\to\eta^{\mu\nu}\), one has \(\xi_0\to0\) and \(\xi_1^2\to\omega^2-\mathbf{k}_\perp^2-m^2\), recovering the standard reduced Green function.

Finally, the phase should not be treated as a harmless prefactor. In the coincidence limit \(z'\to z\), it becomes unity as a multiplicative factor. However, the stress tensor contains derivatives of the Green function. Derivatives with respect to the normal coordinate act on the phase and generate finite terms proportional to \(\xi_0\). These terms will be essential in the next section, because they ensure that the energy-density numerator contains the same Schur-complement structure that controls the spectral part of the reduced Green function.

The discussion in this section is kinematical. It shows that any regular constant quadratic sector admits a reduced Green function whose spectral dependence is governed by the Schur complement \(M^{AB}\). The next step is to show that the energy insertion is governed by the same reduced quadratic structure. This is the part of the derivation that cannot be replaced by diagonalizing the Green function alone.

\section{Energy insertion and the reduced Ward identity}
\label{sec:energy_insertion}

The Casimir energy is obtained from the vacuum expectation value of the energy density, not from the Green function alone. For the quadratic action \eqref{eq:general_quadratic_action}, the canonical stress tensor is
\begin{equation}
\label{eq:general_stress_tensor}
T^\mu{}_{\nu}
=
H^{\mu\alpha}\partial_\alpha\psi\,\partial_\nu\psi
-
\delta^\mu{}_{\nu}\mathcal{L}_2 ,
\end{equation}
where \(\mathcal{L}_2\) is the Lagrangian density appearing in Eq.~\eqref{eq:general_quadratic_action}. Since the plates preserve time-translation invariance, the energy density \(T^0{}_0\) is the relevant Noether density for the global vacuum energy.

For the structural reduction of the global energy, the relevant derivative
part of \(T^0{}_0\) is
\begin{equation}
\label{eq:energy_derivative_part}
H^{0\alpha}\partial_\alpha\psi\,\partial_0\psi .
\end{equation}
In the mixed representation this produces the algebraic numerator
\begin{equation}
\label{eq:unreduced_energy_numerator}
N(q,\kappa)
=
\kappa_0
\left(
H^{0A}\kappa_A
+
H^{03}q
\right),
\end{equation}
with the Fourier convention fixed above. The important point is that the numerator contains the normal momentum \(q\) whenever \(H^{03}\neq0\). Therefore it is sensitive to the same normal-parallel mixing that appears in the reduced Green function.

The reduced numerator is obtained by applying the same shift of the normal momentum that led to Eq.~\eqref{eq:general_square_completion}. Equivalently, at the level of the Green function, it is obtained by allowing the normal derivative in the stress tensor to act on the phase in Eq.~\eqref{eq:general_phase_factor} before taking the coincidence limit. In either language, the reduced normal momentum is replaced by
\begin{equation}
\label{eq:q_reduced_shift}
q
\longrightarrow
-\frac{H^{3B}\kappa_B}{H^{33}} .
\end{equation}
Substituting this into Eq.~\eqref{eq:unreduced_energy_numerator} gives
\begin{equation}
\label{eq:reduced_energy_numerator_1}
N_{\rm red}(\kappa)
=
\kappa_0
\left(
H^{0B}
-
\frac{H^{03}H^{3B}}{H^{33}}
\right)
\kappa_B .
\end{equation}
Using the definition \eqref{eq:schur_complement}, this becomes
\begin{equation}
\label{eq:reduced_energy_numerator}
N_{\rm red}(\kappa)
=
\kappa_0 M^{0B}\kappa_B .
\end{equation}

On the other hand, the reduced bulk quadratic symbol that controls the spectral denominator
of the Green function is
\begin{equation}
\label{eq:general_reduced_inverse}
\Pi_{\rm red}(\kappa)
=
\kappa_A M^{AB}\kappa_B
-
m^2 .
\end{equation}
Since \(M^{AB}\) is symmetric, differentiation with respect to the frequency variable gives
\begin{equation}
\label{eq:freq_derivative_inverse}
\frac{\partial \Pi_{\rm red}}{\partial\kappa_0}
=
2M^{0B}\kappa_B .
\end{equation}
Therefore,
\begin{equation}
\label{eq:general_reduced_ward}
N_{\rm red}(\kappa)
=
\frac12\,\kappa_0
\frac{\partial \Pi_{\rm red}(\kappa)}
{\partial\kappa_0}.
\end{equation}

Equation~\eqref{eq:general_reduced_ward} is the structural identity
behind the effective-metric Casimir formula. After Fourier reduction
parallel to the plates, it shows that the energy numerator
$N_{\rm red}$ is obtained from the frequency derivative of the same
Schur-complement quadratic form $\Pi_{\rm red}$ that fixes
$\xi_1$, and hence the spectral denominator of the reduced boundary
Green function. In this sense, it is the reduced counterpart of the Ward
identity associated with time translations. The plates modify the allowed
normal modes through the boundary Green function, but they do not alter
this bulk relation between the quadratic operator and the energy
insertion.

This point is essential because diagonalizing $\Pi_{\rm red}$ alone
only puts the spectral denominator into a standard form; it does not by
itself guarantee that the energy-density insertion transforms into the
Lorentz-invariant numerator. The missing guarantee is precisely
Eq.~\eqref{eq:general_reduced_ward}. The phase in the reduced Green
function is part of this mechanism: when derivatives act on it, they
supply the $H^{3A}$-dependent terms required for the numerator to
contain the same Schur complement $M^{AB}$. Thus the determinant factor
and the effective plate separation arise from a common reduced quadratic
structure, not merely from diagonalizing the reduced Green function.

The term proportional to \(-\delta^\mu{}_\nu\mathcal{L}_2\) in
Eq.~\eqref{eq:general_stress_tensor} is required in the full local
stress tensor. In the global energy calculation, however, its integrated
contribution is either zero after the vacuum subtraction or reduces to
an \(L\)-independent constant \cite{Milton2001,Escobar2020Local}. It therefore does not contribute to the
plate-separation-dependent Casimir energy, and it drops out of the
pressure obtained by differentiating with respect to \(L\). For this
reason, the structural reduction of the global energy can be analyzed
from the derivative part of the stress tensor alone. The Lagrangian-density term may still be relevant for local quantities, but it does not alter the Schur-complement identity \eqref{eq:general_reduced_ward}, which relates the energy-density numerator to the reduced bulk quadratic symbol.

\section{Effective-metric Casimir formula}
\label{sec:effective_metric_formula}

We now use the preceding identity to obtain the general effective-metric form of the Casimir energy. The result applies to any regular quadratic sector of the form \eqref{eq:general_quadratic_action}, provided that \(H^{\mu\nu}\) is constant, nondegenerate, and has Lorentzian signature, and provided that the boundary conditions isolate the sector under consideration.

After Wick rotation, the reduced bulk quadratic symbol
\eqref{eq:general_reduced_inverse} becomes a Euclidean quadratic form in
the frequency and the two momenta parallel to the plates. In what
follows we restrict ourselves to regular and stable sectors, continuously
connected to the Lorentz-invariant case, for which this Euclidean
quadratic form is positive definite. This assumption is the Euclidean
counterpart of requiring a well-defined fluctuation spectrum and excludes
degenerate or nonhyperbolic sectors. We denote the Wick-rotated parallel
variables collectively by \(\kappa_E\). Since the quadratic form is
positive definite in the sector under consideration, there exists a real
linear transformation from \(\kappa_E\) to new variables
\(r=(r_0,r_1,r_2)\) such that

\begin{equation}
\label{eq:radial_reduction_general}
\Pi_{{\rm red},E}(\kappa_E)
=
r^2+m^2 .
\end{equation}
This is the precise sense in which the momentum-dependent part of the reduced Green function is mapped to the standard Lorentz-invariant Euclidean form.

The transformation also changes the integration measure. Let \(S_E\) denote
the positive Euclidean matrix associated with the Wick-rotated reduced
quadratic form, so that
\begin{equation}
\label{eq:SE_definition}
\Pi_{{\rm red},E}(\kappa_E)
=
\kappa_E^T S_E \kappa_E + m^2 .
\end{equation}
The linear transformation to the variables \(r\) may be chosen such that
\begin{equation}
\label{eq:r_SE_relation}
r^2=\kappa_E^T S_E\kappa_E ,
\end{equation}
and therefore the integration measure transforms as
\begin{equation}
\label{eq:jacobian_SE}
d^3\kappa_E
=
\frac{d^3r}{\sqrt{\det S_E}} .
\end{equation}
The determinant of \(S_E\) is the Euclidean counterpart of the determinant
of the reduced Schur-complement matrix \(M^{AB}\). Using the block
determinant identity
\begin{equation}
\label{eq:block_determinant_general}
\det H
=
H^{33}\det M ,
\end{equation}
the Jacobian can be expressed in terms of \(H^{33}\) and \(\det H\). The
signs are fixed by requiring continuity with the Lorentz-invariant case,
\(H^{\mu\nu}=\eta^{\mu\nu}\), for which \(H^{33}=-1\) and \(\det H=-1\).
With this convention, the determinant factor entering the energy is the
real positive quantity
\begin{equation}
\label{eq:det_factor_positive}
\sqrt{\frac{H^{33}}{\det H}} .
\end{equation}
We now show that the same relation also fixes the effective separation. The boundary
Green function depends on the normal coordinate through the products
\(\xi_1 z\) and \(\xi_1 L\). After Wick rotation and diagonalization of the reduced quadratic form $\Pi_{{\rm red},E}$,
Eq.~\eqref{eq:general_xi1} shows that the Euclidean normal wave number is
rescaled by $1/\sqrt{-H^{33}}$. Thus the normal dependence is identical to the
Lorentz-invariant one after the replacement
\begin{equation}
\label{eq:normal_rescaling_general}
\widetilde z
=
\frac{z}{\sqrt{-H^{33}}},
\qquad
\widetilde L
=
\frac{L}{\sqrt{-H^{33}}}.
\end{equation}
Thus the plate-dependent part of the Green-function integral has the same functional dependence as in the Lorentz-invariant problem, but evaluated at the effective separation \(\widetilde L\).

It remains to check the numerator. By Eq.~\eqref{eq:general_reduced_ward},
the Euclidean continuation of the energy numerator,
\(N_{{\rm red},E}\), is determined by the same reduced quadratic form
that appears in the spectral denominator,
\(\Pi_{{\rm red},E}\). After the linear transformation to the variables
\(r=(r_0,r_1,r_2)\), where \(r_0\) denotes the Euclidean frequency
component in the diagonalized Lorentz-invariant form, the numerator
\(N_{{\rm red},E}\) becomes a quadratic form in \(r\). It is not generally
equal pointwise to the standard Lorentz-invariant numerator \(r_0^2\).
However, the remaining integrand is a radial function of \(r^2\).
Therefore one may use the angular identity
\begin{equation}
\label{eq:angular_identity_general}
\int d^3r\,r_a r_b\,F(r^2)
=
\frac{\delta_{ab}}{3}
\int d^3r\,r^2F(r^2).
\end{equation}
To see why the coefficient is the Lorentz-invariant one, we denote the
Euclidean reduced numerator by \(N_E\equiv N_{{\rm red},E}\). From
Eq.~\eqref{eq:general_reduced_ward}, it is given by
\begin{equation}
\label{eq:euclidean_reduced_numerator}
\begin{aligned}
N_E(\kappa_E)
&\equiv N_{{\rm red},E}(\kappa_E)
\\
&=
\kappa_{E,0}(S_E\kappa_E)_0
\\
&=
\frac12\,\kappa_{E,0}
\frac{\partial \Pi_{{\rm red},E}}{\partial \kappa_{E,0}} .
\end{aligned}
\end{equation}
With \(r=S_E^{1/2}\kappa_E\), this becomes
\begin{equation}
N_E(r)
=
\left(S_E^{-1/2}r\right)_0
\left(S_E^{1/2}r\right)_0 .
\end{equation}
Equivalently,
\begin{equation}
N_E(r)=A_{ab}r_a r_b,
\qquad
A_{ab}\equiv
(S_E^{-1/2})_{0a}(S_E^{1/2})_{0b},
\end{equation}
where \(a,b=0,1,2\). Since \(S_E\) is symmetric and the symmetric square
root is used, one has
\begin{equation}
A_{ab}\delta_{ab}
=
(S_E^{-1/2})_{0a}(S_E^{1/2})_{0b}\delta_{ab}
=
1 .
\end{equation}
Using Eq.~\eqref{eq:angular_identity_general}, the angular average gives
\begin{equation}
\begin{aligned}
\int d^3r\,N_E(r)F(r^2)
&=
A_{ab}\int d^3r\,r_a r_bF(r^2)
\\
&=
\frac{1}{3}A_{ab}\delta_{ab}
\int d^3r\,r^2F(r^2)
\\
&=
\frac{1}{3}
\int d^3r\,r^2F(r^2).
\end{aligned}
\end{equation} 
This is precisely the angular reduction obtained in the
Lorentz-invariant calculation with numerator \(r_0^2\). Therefore the
factor \(1/3\) is not an additional prefactor in the final result; it is
already part of the standard radial integral defining \(E_0\). The only
remaining changes are the Jacobian of the momentum transformation and
the rescaling of the plate separation.

This step is where the Ward-type identity \eqref{eq:general_reduced_ward} is essential. A generic quadratic numerator unrelated to \(\Pi_{\rm red}\) would not reduce in this way.

Combining the radial form of the reduced bulk quadratic symbol, the angular reduction of the numerator, the Jacobian of the momentum transformation, and the rescaling \eqref{eq:normal_rescaling_general}, one obtains the global Casimir energy
\begin{equation}
\label{eq:general_effective_energy}
E_C^{(H)}(L)
=
\sqrt{\frac{H^{33}}{\det H}}\,
E_0\left(
\frac{L}{\sqrt{-H^{33}}}
\right).
\end{equation}
Here \(E_0\) is the Lorentz-invariant scalar Casimir energy for the same boundary condition and mass, evaluated at the rescaled plate separation. For plates normal to a generic spatial direction \(n\), one replaces \(H^{33}\) by \(H^{nn}\).

The corresponding pressure follows either from the normal-normal component of the stress tensor or by differentiating the energy with respect to \(L\), keeping the background tensor fixed. It is
\begin{equation}
\label{eq:general_effective_pressure}
P_C^{(H)}(L)
=
\frac{1}{\sqrt{-\det H}}\,
P_0\left(
\frac{L}{\sqrt{-H^{33}}}
\right).
\end{equation}
Equations~\eqref{eq:general_effective_energy} and
\eqref{eq:general_effective_pressure} reproduce the scalar
Lorentz-violating result of Ref.~\cite{Escobar2020PLB} by setting
\begin{equation}
H^{\mu\nu}=h^{\mu\nu}.
\end{equation}
With this identification, Eq.~\eqref{eq:general_effective_energy}
coincides with Eq.~\eqref{eq:plb_formula_intro}, while
Eq.~\eqref{eq:general_effective_pressure} coincides with
Eq.~\eqref{eq:plb_pressure_intro}. The derivation given here shows
that the result is not tied to the interpretation of \(h^{\mu\nu}\)
as an externally prescribed Lorentz-violating tensor. It follows
from the common quadratic structure of the spectral part of the reduced Green function and
the energy-density insertion.

In applying these formulas, \(H^{\mu\nu}\) denotes the tensor that appears in the quadratic fluctuation operator with the normalization inherited from the underlying linearized action. Thus, when an effective metric is obtained from an optical dispersion relation, its representative must be the one fixed by the corresponding quadratic fluctuation problem, not an arbitrary conformal representative of the same cone. For the massless electromagnetic branches considered below, the final Casimir factors are also checked independently by direct mode summation, which fixes the relevant normalization operationally.

In the rest of the paper we use Eqs.~\eqref{eq:general_effective_energy} and \eqref{eq:general_effective_pressure} as an effective-metric prescription. Whenever the linearization of a nonlinear theory around a constant background produces a regular quadratic branch governed by a constant effective tensor \(H^{\mu\nu}\), the corresponding Casimir contribution is obtained from the formulas above. Nonlinear scalar theories provide a direct realization of this mechanism. Nonlinear electrodynamics provides a more restrictive test, because the linearized electromagnetic sector first decomposes into optical branches; only after this branch decomposition can the effective-metric formula be applied branch by branch.

\section{Nonlinear scalar theories}
\label{sec:scalar}

We first apply the general result of Secs.~\ref{sec:general_quadratic}--\ref{sec:effective_metric_formula} to nonlinear scalar theories. This example is useful for two reasons. First, it shows explicitly how the tensor \(H^{\mu\nu}\) entering the effective-metric Casimir formula can arise dynamically from a nonlinear field theory rather than being introduced as an external Lorentz-violating background. Second, it separates two different effects that are sometimes conflated: nonlinearities in the potential modify the effective mass of the fluctuation, whereas nonlinearities in the kinetic sector modify the effective geometry.

Consider a scalar theory of the form
\begin{equation}
\label{eq:px_scalar_lagrangian}
\mathcal{L}
=
P(X)-U(\Phi),
\qquad
X=
\frac12\eta^{\mu\nu}\partial_\mu\Phi\,\partial_\nu\Phi ,
\end{equation}
where \(\eta_{\mu\nu}=\mathrm{diag}(+,-,-,-)\). We expand the field around a classical background,
\begin{equation}
\label{eq:scalar_background_expansion}
\Phi(x)=\bar\Phi(x)+\varphi(x),
\end{equation}
and assume that the background has a constant gradient,
\begin{equation}
\label{eq:constant_gradient_background}
\partial_\mu\bar\Phi=b_\mu=\mathrm{constant},
\qquad
\bar X=
\frac12 b_\mu b^\mu .
\end{equation}
This assumption is essential for the direct use of the effective-metric formula. It ensures that the coefficients in the quadratic fluctuation operator are constant. If the background were spatially dependent, the reduced Green-function problem would no longer be governed by a constant quadratic symbol, and the formulas derived above would not apply without further analysis.

Expanding \(P(X)\) to second order in \(\varphi\), one finds
\begin{equation}
\label{eq:scalar_deltaX}
X
=
\bar X
+
b^\mu\partial_\mu\varphi
+
\frac12
\eta^{\mu\nu}\partial_\mu\varphi\,\partial_\nu\varphi .
\end{equation}
The term linear in \(\varphi\) vanishes once the background satisfies the classical field equation, or equivalently may be removed from the quadratic fluctuation problem. The kinetic part of the quadratic action is then governed by the Hessian of the Lagrangian with respect to the field gradients:
\begin{equation}
\label{eq:scalar_effective_metric}
H_{\rm sc}^{\mu\nu}
=
P_X(\bar X)\eta^{\mu\nu}
+
P_{XX}(\bar X)b^\mu b^\nu .
\end{equation}
Thus the quadratic action for the fluctuation takes the form
\begin{equation}
\label{eq:scalar_quadratic_action}
S_2[\varphi]
=
\frac12
\int d^4x
\left[
H_{\rm sc}^{\mu\nu}\partial_\mu\varphi\,\partial_\nu\varphi
-
m_{\rm eff}^2\varphi^2
\right],
\end{equation}
where
\begin{equation}
\label{eq:scalar_effective_mass}
m_{\rm eff}^2
=
U''(\bar\Phi)
\end{equation}
whenever the background value of \(U''\) is constant. More generally, possible additional background-dependent terms must also be constant for the present construction to apply.

Equation~\eqref{eq:scalar_quadratic_action} has precisely the form of the general quadratic sector discussed in Sec.~\ref{sec:general_quadratic}, with
\begin{equation}
\label{eq:H_scalar_identification}
H^{\mu\nu}=H_{\rm sc}^{\mu\nu}.
\end{equation}
The origin of the effective tensor is now transparent. A purely nonlinear potential \(U(\Phi)\) changes only the mass term of the fluctuation, through \(U''(\bar\Phi)\). It does not modify the tensor multiplying the derivatives if the kinetic term is canonical. By contrast, a nonlinear dependence on \(X\) changes the Hessian with respect to field gradients and therefore generates a nontrivial effective kinetic geometry. This is the scalar-field analogue of the effective backgrounds considered phenomenologically in Lorentz-violating Casimir studies.

Provided that \(H_{\rm sc}^{\mu\nu}\) is nondegenerate, has Lorentzian signature, and satisfies \(H_{\rm sc}^{nn}<0\) for the plate normal \(n\), the Casimir energy follows directly from Eq.~\eqref{eq:general_effective_energy}:
\begin{equation}
\label{eq:scalar_effective_casimir}
E_C^{\rm sc}(L)
=
\sqrt{
\frac{H_{\rm sc}^{nn}}{\det H_{\rm sc}}
}
\,
E_0\left(
\frac{L}{\sqrt{-H_{\rm sc}^{nn}}};m_{\rm eff}
\right).
\end{equation}
Here \(E_0(L;m_{\rm eff})\) denotes the Lorentz-invariant scalar Casimir energy for a field of mass \(m_{\rm eff}\), evaluated at the indicated separation. The corresponding pressure is
\begin{equation}
\label{eq:scalar_effective_pressure}
P_C^{\rm sc}(L)
=
\frac{1}{\sqrt{-\det H_{\rm sc}}}
\,
P_0\left(
\frac{L}{\sqrt{-H_{\rm sc}^{nn}}};m_{\rm eff}
\right),
\end{equation}
with the background held fixed when differentiating with respect to \(L\).

A simple illustration is obtained by taking
\begin{equation}
\label{eq:scalar_example_PX}
P(X)=X+\frac{\beta}{2\Lambda^4}X^2 .
\end{equation}

Here \(\Lambda\) is the energy scale suppressing the higher-derivative effective interaction, while \(\beta\) is a dimensionless parameter that fixes the sign and strength of the leading nonlinear kinetic correction. The canonical scalar theory is recovered in the limit \(\beta\to0\) or, equivalently, \(\Lambda\to\infty\) at fixed background gradient. For this choice of \(P(X)\), the effective kinetic tensor becomes
\begin{equation}
\label{eq:scalar_example_H}
H_{\rm sc}^{\mu\nu}
=
\left(
1+\frac{\beta\bar X}{\Lambda^4}
\right)\eta^{\mu\nu}
+
\frac{\beta}{\Lambda^4}b^\mu b^\nu .
\end{equation}
For \(\beta=0\), the kinetic tensor reduces to the Minkowski metric and the usual Lorentz-invariant scalar Casimir result is recovered. For a nonzero constant gradient \(b^\mu\), the fluctuation sees a preferred direction, and the Casimir energy depends on the orientation of this direction relative to the plates through \(H_{\rm sc}^{nn}\) and \(\det H_{\rm sc}\). This example shows explicitly how an effective Lorentz-violating Casimir geometry can emerge from a Lorentz-invariant nonlinear scalar theory expanded around a nontrivial background.

The scalar case is therefore a direct realization of the general construction. Its simplicity comes from the fact that the quadratic fluctuation operator is already of effective-metric form. Nonlinear electrodynamics is less immediate: the linearized electromagnetic theory first produces a constitutive tensor and an optical branch structure. The next sections show how, in regular \(\mathcal{L}(\mathcal F)\) theories around a constant magnetic background, the same effective-metric prescription applies after the physical branches have been identified.

\section{Linearized \texorpdfstring{\(\mathcal{L}(\mathcal{F})\)}{L(F)} electrodynamics and optical branches}
\label{sec:nled_linearization}

We now turn to nonlinear electrodynamics in the standard Lagrangian formulation. The purpose of this section is to identify the physical fluctuation branches generated by a constant magnetic background and to determine the effective optical tensors that will enter the branchwise Casimir calculation. This step is essential because, unlike the nonlinear scalar example of Sec.~\ref{sec:scalar}, a nonlinear electromagnetic theory does not generically produce a single second-rank effective metric at the level of the full vector field. Rather, the quadratic expansion defines an effective constitutive law, and the optical metrics arise only after the physical polarization problem has been solved. This is the standard setting in which nonlinear electrodynamics exhibits modified light cones and, in more general models, birefringence \cite{BialynickaBirula1970,Boillat1970,Novello2000,ObukhovRubilar2002,HehlObukhovRubilar2002,DeMelo2014,RussoTownsend2023}.

We consider theories of the form
\begin{equation}
\label{eq:nled_lagrangian}
\mathcal{L}=\mathcal{L}(\mathcal{F}),
\qquad
\mathcal{F}
=
\frac14F_{\mu\nu}F^{\mu\nu}
=
\frac12(\mathbf{B}^2-\mathbf{E}^2),
\end{equation}
with metric convention \(\eta_{\mu\nu}=\mathrm{diag}(+,-,-,-)\). Maxwell electrodynamics corresponds to
\begin{equation}
\label{eq:maxwell_lagrangian}
\mathcal{L}_{\rm M}=-\mathcal{F}.
\end{equation}
We denote derivatives of the Lagrangian evaluated on the background by
\begin{equation}
\label{eq:nled_derivatives}
\mathcal{L}_{\mathcal{F}}
=
\left.
\frac{d\mathcal{L}}{d\mathcal{F}}
\right|_{\bar{\mathcal{F}}},
\qquad
\mathcal{L}_{\mathcal{F}\mathcal{F}}
=
\left.
\frac{d^2\mathcal{L}}{d\mathcal{F}^2}
\right|_{\bar{\mathcal{F}}}.
\end{equation}

The background is taken to be purely magnetic and constant:
\begin{equation}
\label{eq:magnetic_background}
\bar{\mathbf{E}}=0,
\qquad
\bar{\mathbf{B}}=B_0\hat{\mathbf{b}},
\qquad
\bar{\mathcal{F}}=\frac12B_0^2 .
\end{equation}
We write the fluctuating fields as
\begin{equation}
\label{eq:em_fluctuations}
\mathbf{E}=\mathbf{e},
\qquad
\mathbf{B}=\bar{\mathbf{B}}+\mathbf{b}.
\end{equation}
Then
\begin{equation}
\label{eq:F_expansion}
\mathcal{F}
=
\bar{\mathcal{F}}
+
\delta\mathcal{F}^{(1)}
+
\delta\mathcal{F}^{(2)},
\end{equation}
where
\begin{equation}
\label{eq:F_expansion_terms}
\delta\mathcal{F}^{(1)}
=
\bar{\mathbf{B}}\cdot\mathbf{b},
\qquad
\delta\mathcal{F}^{(2)}
=
\frac12(\mathbf{b}^2-\mathbf{e}^2).
\end{equation}
Keeping terms up to second order in the fluctuations gives
\begin{equation}
\label{eq:nled_quadratic_lagrangian}
\mathcal{L}_2
=
\frac12\mathcal{L}_{\mathcal{F}}(\mathbf{b}^2-\mathbf{e}^2)
+
\frac12\mathcal{L}_{\mathcal{F}\mathcal{F}}
(\bar{\mathbf{B}}\cdot\mathbf{b})^2 .
\end{equation}
This quadratic Lagrangian describes a linear anisotropic medium induced by the nonlinear electromagnetic response around the constant magnetic background.

The corresponding linearized constitutive fields are
\begin{equation}
\label{eq:D_H_definitions}
\mathbf{D}
=
\frac{\partial\mathcal{L}_2}{\partial\mathbf{e}},
\qquad
\mathbf{H}
=
-\frac{\partial\mathcal{L}_2}{\partial\mathbf{b}} .
\end{equation}
Thus
\begin{equation}
\label{eq:D_linearized}
\mathbf{D}
=
-\mathcal{L}_{\mathcal{F}}\mathbf{e},
\end{equation}
and
\begin{equation}
\label{eq:H_linearized}
\mathbf{H}
=
-\mathcal{L}_{\mathcal{F}}\mathbf{b}
-
\mathcal{L}_{\mathcal{F}\mathcal{F}}
(\bar{\mathbf{B}}\cdot\mathbf{b})\bar{\mathbf{B}} .
\end{equation}
The electric response is isotropic, whereas the magnetic response distinguishes the component of \(\mathbf{b}\) parallel to the background magnetic field.

The linearized Maxwell equations are
\begin{align}
\label{eq:linearized_maxwell}
\nabla\cdot\mathbf{b}&=0,
&
\nabla\times\mathbf{e}&=-\partial_t\mathbf{b},
\\
\nabla\cdot\mathbf{D}&=0,
&
\nabla\times\mathbf{H}&=\partial_t\mathbf{D}.
\end{align}
For plane waves proportional to \(e^{i(\mathbf{k}\cdot\mathbf{x}-\omega t)}\), these imply
\begin{equation}
\label{eq:plane_wave_relations}
\mathbf{k}\cdot\mathbf{e}=0,
\qquad
\mathbf{b}=\frac{\mathbf{k}\times\mathbf{e}}{\omega},
\end{equation}
provided \(\mathcal{L}_{\mathcal{F}}\neq0\). Substitution of Eqs.~\eqref{eq:D_linearized} and \eqref{eq:H_linearized} into the Ampere equation gives the polarization equation
\begin{equation}
\label{eq:polarization_equation}
(\omega^2-\mathbf{k}^2)\mathbf{e}
+
\frac{\mathcal{L}_{\mathcal{F}\mathcal{F}}}{\mathcal{L}_{\mathcal{F}}}
\left[
\bar{\mathbf{B}}\cdot(\mathbf{k}\times\mathbf{e})
\right]
(\mathbf{k}\times\bar{\mathbf{B}})
=0 .
\end{equation}
Since \(\mathbf{k}\cdot\mathbf{e}=0\), this equation acts on the two-dimensional space of physical transverse polarizations.

Let
\begin{equation}
\label{eq:u_vector}
\mathbf{u}
=
\mathbf{k}\times\bar{\mathbf{B}} .
\end{equation}
One physical polarization is orthogonal to \(\mathbf{u}\). For this mode the second term in Eq.~\eqref{eq:polarization_equation} vanishes, and one obtains the ordinary branch
\begin{equation}
\label{eq:ordinary_branch}
\omega_{\rm ord}^2=\mathbf{k}^2 .
\end{equation}
The complementary physical polarization is parallel to \(\mathbf{u}\). Since \(|\mathbf{k}\times\bar{\mathbf{B}}|^2=B_0^2k_\perp^2\), where
\begin{equation}
\label{eq:k_parallel_perp}
k_\parallel=\mathbf{k}\cdot\hat{\mathbf{b}},
\qquad
k_\perp^2=\mathbf{k}^2-k_\parallel^2,
\end{equation}
this mode satisfies the extraordinary dispersion relation
\begin{equation}
\label{eq:extraordinary_branch}
\omega_{\rm ext}^2
=
k_\parallel^2+\alpha k_\perp^2,
\end{equation}
with
\begin{equation}
\label{eq:alpha_definition}
\alpha
=
1+
\frac{B_0^2\mathcal{L}_{\mathcal{F}\mathcal{F}}}
{\mathcal{L}_{\mathcal{F}}}
=
\frac{
\mathcal{L}_{\mathcal{F}}
+
2\bar{\mathcal{F}}\mathcal{L}_{\mathcal{F}\mathcal{F}}
}
{\mathcal{L}_{\mathcal{F}}}.
\end{equation}
For Maxwell theory, \(\mathcal{L}_{\mathcal{F}}=-1\) and \(\mathcal{L}_{\mathcal{F}\mathcal{F}}=0\), so \(\alpha=1\) and the two branches coincide, as expected.

The regular sector considered in the rest of the paper is defined by
\begin{equation}
\label{eq:nled_regularity_conditions}
\mathcal{L}_{\mathcal{F}}\neq0,
\qquad
\mathcal{L}_{\mathcal{F}}
+
2\bar{\mathcal{F}}\mathcal{L}_{\mathcal{F}\mathcal{F}}
\neq0,
\qquad
\alpha>0 .
\end{equation}
For the Maxwell-connected stable branch, this is naturally supplemented by
\begin{equation}
\label{eq:nled_stability_conditions}
-\mathcal{L}_{\mathcal{F}}>0,
\qquad
-\mathcal{L}_{\mathcal{F}}
-
2\bar{\mathcal{F}}\mathcal{L}_{\mathcal{F}\mathcal{F}}>0 .
\end{equation}
These conditions have a precise role in the present analysis. They ensure
a positive electric response and a nondegenerate magnetic response in the
direction selected by the background field. Equivalently, in the
Maxwell-connected branch they imply \(\alpha>0\), and therefore guarantee
that the extraordinary branch is regular and hyperbolic. They should not,
however, be confused with the full causality requirements usually imposed
in nonlinear electrodynamics. If one further requires the extraordinary
optical cone to lie inside or on the Minkowski light cone, then the
magnetic-background dispersion relation
\(\omega_{\rm ext}^2=k_\parallel^2+\alpha k_\perp^2\) implies the
additional condition \(0<\alpha\leq1\). In the sector with
\(\mathcal{L}_{\mathcal F}<0\), this subluminality condition is equivalent
to \(\mathcal{L}_{\mathcal F\mathcal F}\ge0\). This distinction between
regular hyperbolic propagation and the stronger requirement of causal
propagation relative to the background spacetime is consistent with the
general analyses of optical metrics and energy conditions in nonlinear
electrodynamics \cite{Schellstede2016Causality,RussoTownsend2024Causality}.
Since the Casimir calculation below only requires the regular branch
structure and a real mode spectrum, we impose \(\alpha>0\) throughout,
while noting explicitly when stronger causal assumptions would be needed.

The ordinary branch is described by
\begin{equation}
\label{eq:h_ord}
h_{\rm ord}^{\mu\nu}=\eta^{\mu\nu}.
\end{equation}
The extraordinary branch can be represented by the optical tensor
\begin{align}
\label{eq:h_ext_general}
h_{\rm ext}^{00}&=1,\qquad
h_{\rm ext}^{0i}=0,
\\ 
h_{\rm ext}^{ij}
&=
-\left[
\alpha\delta^{ij}
+
(1-\alpha)\hat b^i\hat b^j
\right]. \nonumber
\end{align}
Here $\hat{\mathbf b}=\bar{\mathbf B}/B_0$ is the unit vector along the constant magnetic background.
 This tensor reproduces Eq.~\eqref{eq:extraordinary_branch} through
\(h_{\rm ext}^{\mu\nu}k_\mu k_\nu=0\). The representatives in Eq.~\eqref{eq:h_ext_general} are fixed by the frequency spectra obtained from the linearized field equations. They are not arbitrary conformal representatives of the optical cone.

\section{Direct mode summation and boundary conditions}
\label{sec:mode_sum}

We now compute the parallel-plate Casimir energy directly from the ordinary and extraordinary frequency spectra. This calculation provides an independent check of the branchwise effective-metric prescription developed above.

We consider two perfectly conducting plates located at \(z=0\) and \(z=L\), with normal vector \(\hat{\mathbf{z}}\). The constant magnetic background is assumed to be maintained by external sources and is not varied in the fluctuation problem. The perfect-conductor boundary conditions are imposed on the fluctuating radiation fields,
\begin{equation}
\label{eq:perfect_conductor_bc}
\hat{\mathbf{z}}\times\mathbf{e}=0,
\qquad
\hat{\mathbf{z}}\cdot\mathbf{b}=0
\qquad
(z=0,L).
\end{equation}
We work directly with the physical fields \(\mathbf{e}\) and \(\mathbf{b}\). Thus no gauge modes are introduced. The two branches derived in Sec.~\ref{sec:nled_linearization} represent the two physical electromagnetic polarizations: one ordinary branch and one extraordinary branch.

We denote by
\begin{equation}
\label{eq:E_one_definition}
E_0^{(1)}(L)
=
-\frac{\pi^2}{1440L^3}
\end{equation}
the renormalized \(L\)-dependent Casimir energy per unit area of a single
massless physical branch with normal spectrum \(k_z=n\pi/L\). This
notation includes the standard TE/TM mode-counting convention for one
physical polarization. The two Maxwell polarizations therefore give
\begin{equation}
\label{eq:maxwell_energy_twice_scalar}
E_{\rm Maxwell}(L)
=
2E_0^{(1)}(L)
=
-\frac{\pi^2}{720L^3}.
\end{equation}

\subsection{Compatibility with the conducting boundary conditions}
\label{subsec:boundary_conditions}

Before performing the mode sums, we verify that the branch decomposition is compatible with Eq.~\eqref{eq:perfect_conductor_bc} for the two orientations considered below.

First, take the magnetic background normal to the plates,
\begin{equation}
\label{eq:B_normal}
\bar{\mathbf{B}}=B_0\hat{\mathbf{z}} .
\end{equation}
Then the extraordinary polarization satisfies
\begin{equation}
\label{eq:extra_pol_normal}
\mathbf{e}_{\rm ext}
\parallel
\mathbf{k}\times\hat{\mathbf{z}}
=
(k_y,-k_x,0).
\end{equation}
It is purely tangential to the plates. A standing wave with tangential components proportional to \(\sin(n\pi z/L)\) therefore satisfies \(\hat{\mathbf{z}}\times\mathbf{e}=0\) at \(z=0,L\). Moreover,
\begin{equation}
\label{eq:bz_normal_case}
b_z
=
\frac{1}{\omega}(k_x e_y-k_y e_x)
\end{equation}
is proportional to the same sine factor and also vanishes at the plates. Hence the extraordinary branch is compatible with the standard normal quantization
\begin{equation}
\label{eq:kz_quantization}
k_z=\frac{n\pi}{L}.
\end{equation}

Second, take the magnetic background parallel to the plates, for definiteness
\begin{equation}
\label{eq:B_parallel}
\bar{\mathbf{B}}=B_0\hat{\mathbf{x}} .
\end{equation}
The extraordinary polarization is
\begin{equation}
\label{eq:extra_pol_parallel}
\mathbf{e}_{\rm ext}
\parallel
\mathbf{k}\times\hat{\mathbf{x}}
=
(0,k_z,-k_y).
\end{equation}
A standing wave may be chosen with
\begin{equation}
\label{eq:parallel_standing_wave}
e_y\propto \sin\left(\frac{n\pi z}{L}\right),
\qquad
e_z\propto \cos\left(\frac{n\pi z}{L}\right).
\end{equation}
Since \(e_x=0\), the tangential electric field vanishes at the plates. In addition,
\begin{equation}
\label{eq:bz_parallel_case}
b_z=\frac{1}{\omega}(k_xe_y-k_ye_x)
=
\frac{k_x}{\omega}e_y,
\end{equation}
and therefore \(b_z=0\) at \(z=0,L\). The ordinary branch again admits the usual complementary standing-wave construction. Thus, for both orientations considered in this work, the physical
branches can be quantized with \(k_z=n\pi/L\), and no mixing of the two
polarizations is induced by the perfect-conductor boundary conditions.
The ordinary branch is the polarization orthogonal to
\(\mathbf{k}\times\bar{\mathbf B}\). In the two orientations considered
here it can be chosen as the complementary TE/TM standing wave satisfying
Eq.~\eqref{eq:perfect_conductor_bc}, and it reduces continuously to the
second Maxwell polarization when \(\alpha=1\).

\subsection{Magnetic background normal to the plates}

For \(\bar{\mathbf{B}}=B_0\hat{\mathbf{z}}\), one has
\begin{equation}
\label{eq:normal_k_decomp}
k_\parallel=k_z,
\qquad
k_\perp^2=k_x^2+k_y^2 .
\end{equation}
The extraordinary frequency is therefore
\begin{equation}
\label{eq:omega_ext_normal}
\omega_{{\rm ext},n}^2
=
\left(\frac{n\pi}{L}\right)^2
+
\alpha(k_x^2+k_y^2).
\end{equation}
The corresponding renormalized energy is obtained from
\begin{equation}
\label{eq:E_ext_normal_sum}
E_{\rm ext}^{\perp}(L)
=
\frac12
\sum_n
\int\frac{d^2\mathbf{k}_\perp}{(2\pi)^2}
\sqrt{
\left(\frac{n\pi}{L}\right)^2
+
\alpha(k_x^2+k_y^2)
}\Bigg|_{\rm ren}.
\end{equation}
The change of variables \(q_x=\sqrt{\alpha}k_x\), \(q_y=\sqrt{\alpha}k_y\) gives \(d^2\mathbf{k}_\perp=d^2\mathbf{q}_\perp/\alpha\), and hence
\begin{equation}
\label{eq:E_ext_normal_result}
E_{\rm ext}^{\perp}(L)
=
\frac{1}{\alpha}E_0^{(1)}(L).
\end{equation}
The ordinary branch contributes \(E_0^{(1)}(L)\). Therefore
\begin{equation}
\label{eq:E_total_normal}
E_C^{\perp}(L)
=
\left(1+\frac1\alpha\right)E_0^{(1)}(L)
=
-\frac{\pi^2}{1440L^3}
\left(1+\frac1\alpha\right).
\end{equation}

\subsection{Magnetic background parallel to the plates}

For \(\bar{\mathbf{B}}=B_0\hat{\mathbf{x}}\), one has
\begin{equation}
\label{eq:parallel_k_decomp}
k_\parallel=k_x,
\qquad
k_\perp^2=k_y^2+k_z^2 .
\end{equation}
The extraordinary frequency is
\begin{equation}
\label{eq:omega_ext_parallel}
\omega_{{\rm ext},n}^2
=
k_x^2
+
\alpha k_y^2
+
\alpha\left(\frac{n\pi}{L}\right)^2 .
\end{equation}
Thus
\begin{equation}
\label{eq:E_ext_parallel_sum}
E_{\rm ext}^{\parallel}(L)
=
\frac12
\sum_n
\int\frac{dk_xdk_y}{(2\pi)^2}
\sqrt{
k_x^2+\alpha k_y^2
+\alpha\left(\frac{n\pi}{L}\right)^2
}\Bigg|_{\rm ren}.
\end{equation}
Setting \(q_y=\sqrt{\alpha}k_y\), the measure becomes \(dk_y=dq_y/\sqrt{\alpha}\), and the normal spectrum is equivalent to that of a branch at the effective separation \(L/\sqrt{\alpha}\). Hence
\begin{equation}
\label{eq:E_ext_parallel_intermediate}
E_{\rm ext}^{\parallel}(L)
=
\frac{1}{\sqrt{\alpha}}
E_0^{(1)}
\left(\frac{L}{\sqrt{\alpha}}\right).
\end{equation}
Since \(E_0^{(1)}(L)\propto L^{-3}\), this gives
\begin{equation}
\label{eq:E_ext_parallel_result}
E_{\rm ext}^{\parallel}(L)
=
\alpha E_0^{(1)}(L).
\end{equation}
Adding the ordinary branch,
\begin{equation}
\label{eq:E_total_parallel}
E_C^{\parallel}(L)
=
(1+\alpha)E_0^{(1)}(L)
=
-\frac{\pi^2}{1440L^3}(1+\alpha).
\end{equation}
Equations~\eqref{eq:E_total_normal} and \eqref{eq:E_total_parallel} reduce to the standard Maxwell result \eqref{eq:maxwell_energy_twice_scalar} when \(\alpha=1\).

\section{Branchwise effective-metric computation}
\label{sec:branchwise}

We now reproduce the results of Sec.~\ref{sec:mode_sum} using the effective-metric formula of Sec.~\ref{sec:effective_metric_formula}. This calculation is not applied to the electromagnetic field as a single scalar sector. Rather, it is applied after the physical electromagnetic fluctuations have been decomposed into their ordinary and extraordinary branches. The representatives of the optical tensors are those fixed by the linearized spectra derived in Sec.~\ref{sec:nled_linearization}, and the direct mode-sum calculation above provides an independent normalization check.

The ordinary branch is governed by
\begin{equation}
\label{eq:H_ord_branchwise}
H_{\rm ord}^{\mu\nu}=\eta^{\mu\nu},
\end{equation}
and therefore contributes
\begin{equation}
\label{eq:E_ord_branchwise}
E_{\rm ord}(L)=E_0^{(1)}(L).
\end{equation}

For a magnetic background normal to the plates,
\(\bar{\mathbf{B}}=B_0\hat{\mathbf{z}}\), the extraordinary dispersion relation is
\begin{equation}
\label{eq:disp_ext_normal_branchwise}
\omega_{\rm ext}^2
=
k_z^2+\alpha(k_x^2+k_y^2).
\end{equation}
The corresponding effective tensor is
\begin{equation}
\label{eq:H_ext_normal}
H_{\rm ext}^{\mu\nu}
=
\mathrm{diag}(1,-\alpha,-\alpha,-1).
\end{equation}
Thus
\begin{equation}
\label{eq:H_ext_normal_det}
H_{\rm ext}^{zz}=-1,
\qquad
\det H_{\rm ext}=-\alpha^2 .
\end{equation}
Using Eq.~\eqref{eq:general_effective_energy}, one obtains
\begin{equation}
\label{eq:E_ext_normal_branchwise}
E_{\rm ext}^{\perp}(L)
=
\sqrt{
\frac{-1}{-\alpha^2}
}
E_0^{(1)}(L)
=
\frac1\alpha E_0^{(1)}(L),
\end{equation}
in agreement with Eq.~\eqref{eq:E_ext_normal_result}.

For a magnetic background parallel to the plates,
\(\bar{\mathbf{B}}=B_0\hat{\mathbf{x}}\), the extraordinary dispersion relation is
\begin{equation}
\label{eq:disp_ext_parallel_branchwise}
\omega_{\rm ext}^2
=
k_x^2+\alpha k_y^2+\alpha k_z^2.
\end{equation}
The effective tensor is
\begin{equation}
\label{eq:H_ext_parallel}
H_{\rm ext}^{\mu\nu}
=
\mathrm{diag}(1,-1,-\alpha,-\alpha).
\end{equation}
Therefore
\begin{equation}
\label{eq:H_ext_parallel_det}
H_{\rm ext}^{zz}=-\alpha,
\qquad
\det H_{\rm ext}=-\alpha^2 .
\end{equation}
The effective-metric formula gives
\begin{equation}
\label{eq:E_ext_parallel_branchwise}
E_{\rm ext}^{\parallel}(L)
=
\sqrt{
\frac{-\alpha}{-\alpha^2}
}
E_0^{(1)}
\left(\frac{L}{\sqrt{\alpha}}\right)
=
\frac{1}{\sqrt{\alpha}}
E_0^{(1)}
\left(\frac{L}{\sqrt{\alpha}}\right).
\end{equation}
Using again \(E_0^{(1)}(L)\propto L^{-3}\), this becomes
\begin{equation}
\label{eq:E_ext_parallel_branchwise_final}
E_{\rm ext}^{\parallel}(L)
=
\alpha E_0^{(1)}(L),
\end{equation}
which agrees with Eq.~\eqref{eq:E_ext_parallel_result}.

The comparison can be summarized as

\begin{equation}
\label{eq:mode_sum_branchwise_summary}
\renewcommand{\arraystretch}{1.35}
\setlength{\arraycolsep}{3.5pt}
\begin{array}{|c|c|c|}
\hline
\text{orientation}
&
\text{direct mode sum}
&
\begin{array}{c}
\text{branchwise}\\
\text{effective metric}
\end{array}
\\
\hline
\bar{\mathbf{B}}\perp \text{ plates}
&
E_{\rm ext}^{\perp}=\alpha^{-1}E_0^{(1)}(L)
&
E_{\rm ext}^{\perp}=\alpha^{-1}E_0^{(1)}(L)
\\
\hline
\bar{\mathbf{B}}\parallel \text{ plates}
&
E_{\rm ext}^{\parallel}=\alpha E_0^{(1)}(L)
&
E_{\rm ext}^{\parallel}=\alpha E_0^{(1)}(L)
\\
\hline
\end{array}
\end{equation}

Thus, for regular \(\mathcal{L}(\mathcal{F})\) electrodynamics around a constant magnetic background, the direct electromagnetic mode sum agrees exactly with the effective-metric formula applied branch by branch.

The corresponding pressures follow by differentiating with respect to \(L\), keeping the magnetic background fixed:
\begin{align}
\label{eq:pressure_normal_parallel}
P_C^{\perp}(L)
&=
-\frac{\pi^2}{480L^4}
\left(1+\frac1\alpha\right),
\\
P_C^{\parallel}(L)
&=
-\frac{\pi^2}{480L^4}
(1+\alpha). \label{eq:pressure_normal_parallel1}
\end{align}
Both reduce to the standard electromagnetic pressure,
\(-\pi^2/(240L^4)\), in the Maxwell limit \(\alpha=1\).

The equivalence established in this section has a controlled scope. It assumes a constant magnetic background, the Lagrangian class \(\mathcal{L}(\mathcal{F})\), regular optical sectors with \(\alpha>0\), and the two orientations for which the perfect-conductor boundary conditions preserve the branch decomposition. Oblique magnetic backgrounds, general \(\mathcal{L}(\mathcal{F},\mathcal{G})\) theories, dispersive material boundaries, and degenerate sectors in which the effective constitutive tensor loses rank require separate analyses. In such cases one should not assume a branchwise scalar reduction without verifying the full mode structure and the boundary conditions.


\section{Weakly nonlinear example}
\label{sec:example}

The preceding results are independent of a specific nonlinear model, provided the theory belongs to the regular \(\mathcal{L}(\mathcal{F})\) class and the background is constant. It is nevertheless useful to display the size and sign of the effect in a simple weakly nonlinear example. We consider
\begin{equation}
\label{eq:nled_example_lagrangian}
\mathcal{L}(\mathcal{F})
=
-\mathcal{F}
+
\frac{\gamma}{2\Lambda^4}\mathcal{F}^2 ,
\end{equation}
where \(\Lambda\) is the scale suppressing the nonlinear correction and \(\gamma\) is a dimensionless coefficient. The Maxwell limit is recovered either by taking \(\gamma\to0\) or by sending \(\Lambda\to\infty\) at fixed background field.

This example should be viewed as the leading \(\mathcal{F}^2\) correction in a low-energy effective expansion. It is not meant to represent the full Euler--Heisenberg effective action, which also contains dependence on the second electromagnetic invariant and therefore lies in the broader \(\mathcal{L}(\mathcal{F},\mathcal{G})\) class. The purpose of Eq.~\eqref{eq:nled_example_lagrangian} is instead to illustrate, within the controlled \(\mathcal{L}(\mathcal{F})\) sector analyzed above, how a nonlinear magnetic response modifies the Casimir energy.

For the magnetic background \(\bar{\mathcal{F}}=B_0^2/2\), one has
\begin{equation}
\label{eq:nled_example_derivatives}
\mathcal{L}_{\mathcal{F}}
=
-1+\frac{\gamma\bar{\mathcal{F}}}{\Lambda^4},
\qquad
\mathcal{L}_{\mathcal{F}\mathcal{F}}
=
\frac{\gamma}{\Lambda^4}.
\end{equation}
The anisotropy parameter defined in Eq.~\eqref{eq:alpha_definition} is therefore
\begin{equation}
\label{eq:nled_example_alpha}
\alpha
=
\frac{
-1+\frac{3\gamma B_0^2}{2\Lambda^4}
}{
-1+\frac{\gamma B_0^2}{2\Lambda^4}
}.
\end{equation}
The regular Maxwell-connected sector requires \(\mathcal{L}_{\mathcal{F}}<0\) and \(\alpha>0\). For \(\gamma>0\), a sufficient weak-field condition is
\begin{equation}
\label{eq:nled_example_regular_condition}
\frac{\gamma B_0^2}{\Lambda^4}\ll1 .
\end{equation}
In this regime,
\begin{equation}
\label{eq:nled_example_alpha_weak}
\alpha
=
1-
\frac{\gamma B_0^2}{\Lambda^4}
+
\mathcal{O}\!\left(\frac{B_0^4}{\Lambda^8}\right).
\end{equation}

Using Eqs.~\eqref{eq:E_total_normal} and \eqref{eq:E_total_parallel}, the Casimir energies for magnetic backgrounds normal and parallel to the plates become
\begin{align}
\label{eq:nled_example_energy_perp}
E_C^\perp(L)
&=
-\frac{\pi^2}{1440L^3}
\left[
1+\frac1\alpha
\right],
\\
\label{eq:nled_example_energy_parallel}
E_C^\parallel(L)
&=
-\frac{\pi^2}{1440L^3}
\left[
1+\alpha
\right].
\end{align}
Expanding to leading order in \(\gamma B_0^2/\Lambda^4\), one obtains
\begin{align}
\label{eq:nled_example_weak_perp}
E_C^\perp(L)
&=
-\frac{\pi^2}{720L^3}
\left[
1+
\frac12
\frac{\gamma B_0^2}{\Lambda^4}
+
\mathcal{O}\!\left(\frac{B_0^4}{\Lambda^8}\right)
\right],
\\
\label{eq:nled_example_weak_parallel}
E_C^\parallel(L)
&=
-\frac{\pi^2}{720L^3}
\left[
1-
\frac12
\frac{\gamma B_0^2}{\Lambda^4}
+
\mathcal{O}\!\left(\frac{B_0^4}{\Lambda^8}\right)
\right].
\end{align}
Thus the leading nonlinear correction has opposite signs for the two orientations. For \(\gamma>0\), a magnetic background normal to the plates increases the magnitude of the attractive Casimir energy, while a magnetic background parallel to the plates decreases it. Equivalently, the orientation contrast is
\begin{equation}
\label{eq:nled_example_orientation_contrast}
E_C^\perp(L)-E_C^\parallel(L)
=
-\frac{\pi^2}{720L^3}
\frac{\gamma B_0^2}{\Lambda^4}
+
\mathcal{O}\!\left(\frac{B_0^4}{\Lambda^8}\right).
\end{equation}

The corresponding pressures follow from Eqs.~\eqref{eq:pressure_normal_parallel}-\eqref{eq:pressure_normal_parallel1}. In the same weak-field regime,
\begin{align}
\label{eq:nled_example_pressure_perp}
P_C^\perp(L)
&=
-\frac{\pi^2}{240L^4}
\left[
1+
\frac12
\frac{\gamma B_0^2}{\Lambda^4}
+
\mathcal{O}\!\left(\frac{B_0^4}{\Lambda^8}\right)
\right],
\\
\label{eq:nled_example_pressure_parallel}
P_C^\parallel(L)
&=
-\frac{\pi^2}{240L^4}
\left[
1-
\frac12
\frac{\gamma B_0^2}{\Lambda^4}
+
\mathcal{O}\!\left(\frac{B_0^4}{\Lambda^8}\right)
\right].
\end{align}
These expressions make explicit the physical content of the branchwise calculation. The nonlinear electromagnetic response to the magnetic background does not simply rescale the Maxwell result by a universal constant; it induces an anisotropic Casimir response controlled by the orientation of the background relative to the plates.

\section{Conclusions}
\label{sec:conclusions}

We have given a structural derivation of the effective-metric form of the
parallel-plate Casimir energy for regular quadratic fluctuation sectors
with constant coefficients. The determinant prefactor and the rescaling of the plate separation were already known in the Lorentz-violating scalar-field setting, in particular in the result of Ref.~\cite{Escobar2020PLB}. Rather than deriving the Casimir effect from
scratch, our purpose was to identify why that formula has this structure
and under what conditions it can be used as an effective-metric
prescription.
 The key point is that diagonalizing the
spectral denominator of the reduced boundary Green function is not, by
itself, sufficient. The energy-density insertion must transform
consistently as well. For a general constant-coefficient quadratic
sector, this consistency follows from a common Schur-complement
structure: the same reduced quadratic form fixes the shifted normal wave
number entering the boundary Green function and, through the reduced
Ward-type identity associated with time translations, also fixes the
reduced energy numerator.

This analysis clarifies the status of the effective tensor entering the
Casimir formula. The relevant object is not simply an arbitrary
representative of a characteristic cone, but the tensor that appears in
the normalized quadratic fluctuation operator obtained from the
linearized action. This distinction matters in nonlinear theories,
because effective geometries arise from the background-dependent
quadratic response, whereas optical cones alone are insensitive to
conformal rescalings. In the scalar example, this tensor is the Hessian
of the Lagrangian with respect to the field gradients, evaluated on a
constant-gradient background. A nonlinear potential by itself changes
the fluctuation mass, but it does not generate an anisotropic kinetic
geometry.

For nonlinear electrodynamics of the form \(\mathcal{L}(\mathcal{F})\), the effective-metric construction becomes nontrivial because the electromagnetic fluctuation does not reduce to a single scalar sector. The constant magnetic background first induces an anisotropic linear response, and the physical spectrum splits into an ordinary Maxwell branch and an extraordinary optical branch. We showed that, in the regular sector, the ordinary and extraordinary modes provide the two physical electromagnetic polarizations. For magnetic backgrounds normal or parallel to perfectly conducting plates, the boundary conditions preserve this branch decomposition and lead to the standard normal quantization.

The main electromagnetic result is the exact agreement between two independent calculations. Direct summation of the ordinary and extraordinary mode frequencies gives the same Casimir energies as the branchwise application of the effective-metric formula to the corresponding optical tensors. This agreement fixes the normalization of the optical representatives operationally and supports the use of the effective-metric formula for regular branches, rather than for the electromagnetic field as an undecomposed whole.

Several extensions are natural but require separate treatment. Oblique magnetic backgrounds can mix the polarization structure at the boundaries and should be analyzed with the full boundary mode problem. Purely electric backgrounds also lead, at the level of the linearized bulk equations, to an analogous ordinary--extraordinary branch structure, with the anisotropy controlled by the electric response rather than by the magnetic one. However, their treatment in the Casimir geometry requires a separate analysis. In particular, the compatibility of constant electric backgrounds with conducting boundary conditions is more restrictive, and in quantum effective theories electric backgrounds may introduce additional vacuum-instability issues. We therefore leave electric backgrounds beyond the scope of the present work.

General \(\mathcal{L}(\mathcal{F},\mathcal{G})\) theories may produce a richer birefringent structure and cannot be reduced to the single-parameter \(\alpha\) description used here. Dispersive or material boundaries would also modify the mode spectrum and the renormalization procedure. Finally, degenerate sectors, where the constitutive tensor loses rank or an optical metric becomes singular, lie outside the regular quadratic framework developed in this paper. In such cases the number of propagating degrees of freedom and the constraint structure must be analyzed before any effective-metric Casimir formula can be applied.

\section*{Acknowledgments}
The author is grateful to A. Mart\'in-Ruiz for useful comments and suggestions on the manuscript.

\section*{Data Availability}
No data were created or analyzed in this work.

\bigskip

\bibliographystyle{apsrev4-2}
\bibliography{nled_effective_metric_casimir_refs}

@article{PlacidoFlores2026MagneticVacua,
  author = {Pl{\'a}cido-Flores, E. and Linares, Rom{\'a}n and L{\'o}pez, V. and Escobar, C. A.},
  title = {Stable magnetic {Lorentz}-violating vacua in gauge-invariant nonlinear electrodynamics},
  journal = {Eur. Phys. J. C},
  volume = {86},
  number = {6},
  pages = {619},
  year = {2026},
  doi = {10.1140/epjc/s10052-026-15892-w},
  eprint = {2605.03341},
  archivePrefix = {arXiv},
  primaryClass = {hep-th}
}

@article{EscobarRuiz2021Cylinder,
  author = {Escobar-Ruiz, A. M. and Mart{\'i}n-Ruiz, A. and Escobar, C. A. and Linares, R.},
  title = {Scalar {Casimir} effect for a conducting cylinder in a {Lorentz}-violating background},
  journal = {Int. J. Mod. Phys. A},
  volume = {36},
  number = {23},
  pages = {2150168},
  year = {2021},
  doi = {10.1142/S0217751X21501682},
  eprint = {2105.12953},
  archivePrefix = {arXiv},
  primaryClass = {hep-th}
}

@article{CruzBezerraPetrov2019Fermionic,
  author = {Cruz, M. B. and Bezerra de Mello, E. R. and Petrov, A. Yu.},
  title = {Fermionic {Casimir} effect in a field theory model with {Lorentz} symmetry violation},
  journal = {Phys. Rev. D},
  volume = {99},
  number = {8},
  pages = {085012},
  year = {2019},
  doi = {10.1103/PhysRevD.99.085012},
  eprint = {1812.05428},
  archivePrefix = {arXiv},
  primaryClass = {hep-th}
}

@article{BezerraCruz2023ConstantVectors,
  author = {Bezerra de Mello, E. R. and Cruz, M. B.},
  title = {Scalar {Casimir} effects in a {Lorentz} violation scenario induced by the presence of constant vectors},
  journal = {Int. J. Mod. Phys. A},
  volume = {38},
  number = {11},
  pages = {2350062},
  year = {2023},
  doi = {10.1142/S0217751X23500628},
  eprint = {2210.09243},
  archivePrefix = {arXiv},
  primaryClass = {hep-th}
}

@article{DantasMotaBezerra2023HigherDerivatives,
  author = {Dantas, Robson A. and Mota, Herondy F. Santana and Bezerra de Mello, Eug{\^e}nio R.},
  title = {Bosonic {Casimir} Effect in an Aether-like {Lorentz}-Violating Scenario with Higher Order Derivatives},
  journal = {Universe},
  volume = {9},
  number = {5},
  pages = {241},
  year = {2023},
  doi = {10.3390/universe9050241},
  eprint = {2304.04078},
  archivePrefix = {arXiv},
  primaryClass = {hep-th}
}

@article{Schellstede2016Causality,
author = {Schellstede, Gerold O. and Perlick, Volker and L{\"a}mmerzahl, Claus},
title = {On causality in nonlinear vacuum electrodynamics of the {Pleba{\'n}ski} class},
journal = {Annalen Phys.},
volume = {528},
number = {9--10},
pages = {738--749},
year = {2016},
doi = {10.1002/andp.201600124},
eprint = {1604.02545},
archivePrefix = {arXiv},
primaryClass = {gr-qc}
}

@article{RussoTownsend2024Causality,
author = {Russo, Jorge G. and Townsend, Paul K.},
title = {Causality and energy conditions in nonlinear electrodynamics},
journal = {JHEP},
volume = {2024},
number = {6},
pages = {191},
year = {2024},
doi = {10.1007/JHEP06(2024)191},
eprint = {2404.09994},
archivePrefix = {arXiv},
primaryClass = {hep-th}
}

@article{Plunien1986,
author = {Plunien, G. and M\"uller, B. and Greiner, W.},
title = {The Casimir effect},
journal = {Phys. Rept.},
volume = {134},
pages = {87--193},
year = {1986},
doi = {10.1016/0370-1573(86)90020-7}
}

@article{BrownMaclay1969,
author = {Brown, Lowell S. and Maclay, G. Jordan},
title = {Vacuum Stress between Conducting Plates: An Image Solution},
journal = {Phys. Rev.},
volume = {184},
pages = {1272--1279},
year = {1969},
doi = {10.1103/PhysRev.184.1272}
}

@article{Cruz2018,
author = {Cruz, M. B. and Bezerra de Mello, E. R. and Petrov, A. Yu.},
title = {Thermal corrections to the {Casimir} energy in a {Lorentz}-breaking scalar field theory},
journal = {Mod. Phys. Lett. A},
volume = {33},
number = {20},
pages = {1850115},
year = {2018},
doi = {10.1142/S0217732318501158},
eprint = {1803.07446},
archivePrefix = {arXiv},
primaryClass = {hep-th}
}

@article{BialynickaBirula1970,
author = {Bialynicka-Birula, Z. and Bialynicki-Birula, I.},
title = {Nonlinear Effects in Quantum Electrodynamics. Photon Propagation and Photon Splitting in an External Field},
journal = {Phys. Rev. D},
volume = {2},
pages = {2341--2345},
year = {1970},
doi = {10.1103/PhysRevD.2.2341}
}

@article{ObukhovRubilar2002,
author = {Obukhov, Yuri N. and Rubilar, Guillermo F.},
title = {Fresnel analysis of the wave propagation in nonlinear electrodynamics},
journal = {Phys. Rev. D},
volume = {66},
pages = {024042},
year = {2002},
doi = {10.1103/PhysRevD.66.024042},
eprint = {gr-qc/0204028},
archivePrefix = {arXiv}
}

@article{HehlObukhovRubilar2002,
  author = {Hehl, F. W. and Obukhov, Y. N. and Rubilar, G. F.},
  title = {Light propagation in generally covariant electrodynamics and the {Fresnel} equation},
  journal = {Int. J. Mod. Phys. A},
  volume = {17},
  pages = {2695--2700},
  year = {2002},
  doi = {10.1142/S0217751X0201162X},
  eprint = {gr-qc/0203105},
  archivePrefix = {arXiv},
  primaryClass = {gr-qc}
}

@article{Casimir1948,
  author = {Casimir, H. B. G.},
  title = {On the attraction between two perfectly conducting plates},
  journal = {Proc. Kon. Ned. Akad. Wet.},
  volume = {51},
  pages = {793--795},
  year = {1948}
}

@book{Milton2001,
  author = {Milton, K. A.},
  title = {The Casimir Effect: Physical Manifestations of Zero-Point Energy},
  publisher = {World Scientific},
  address = {Singapore},
  year = {2001}
}

@book{Bordag2009,
  author = {Bordag, M. and Klimchitskaya, G. L. and Mohideen, U. and Mostepanenko, V. M.},
  title = {Advances in the Casimir Effect},
  publisher = {Oxford University Press},
  address = {Oxford},
  year = {2009}
}

@article{Cruz2017,
  author = {Cruz, M. B. and Bezerra de Mello, E. R. and Petrov, A. Yu.},
  title = {Casimir effects in {Lorentz}-violating scalar field theory},
  journal = {Phys. Rev. D},
  volume = {96},
  pages = {045019},
  year = {2017},
  doi = {10.1103/PhysRevD.96.045019},
  eprint = {1705.03331},
  archivePrefix = {arXiv},
  primaryClass = {hep-th}
}

@article{MartinRuiz2017LocalEM,
  author = {Mart{\'i}n-Ruiz, A. and Escobar, C. A.},
  title = {Local effects of the quantum vacuum in {Lorentz}-violating electrodynamics},
  journal = {Phys. Rev. D},
  volume = {95},
  number = {3},
  pages = {036011},
  year = {2017},
  doi = {10.1103/PhysRevD.95.036011},
  eprint = {1611.04616},
  archivePrefix = {arXiv},
  primaryClass = {hep-th}
}

@article{Escobar2020Local,
  author = {Escobar, C. A. and Medel, Leonardo and Mart\'in-Ruiz, A.},
  title = {Casimir effect in Lorentz-violating scalar field theory: A local approach},
  journal = {Phys. Rev. D},
  volume = {101},
  pages = {095011},
  year = {2020},
  doi = {10.1103/PhysRevD.101.095011},
  eprint = {2005.00151},
  archivePrefix = {arXiv},
  primaryClass = {hep-th}
}

@article{Escobar2020PLB,
  author = {Escobar, C. A. and Mart\'in-Ruiz, A. and Franca, O. J. and Garcia, Marcos A. G.},
  title = {A non-perturbative approach to the scalar Casimir effect with Lorentz symmetry violation},
  journal = {Phys. Lett. B},
  volume = {807},
  pages = {135567},
  year = {2020},
  doi = {10.1016/j.physletb.2020.135567},
  eprint = {2005.14217},
  archivePrefix = {arXiv},
  primaryClass = {hep-th}
}

@article{MartinRuiz2020Sphere,
  author = {Mart\'in-Ruiz, A. and Escobar, C. A. and Escobar-Ruiz, A. M. and Franca, O. J.},
  title = {Lorentz violating scalar Casimir effect for a D-dimensional sphere},
  journal = {Phys. Rev. D},
  volume = {102},
  pages = {015027},
  year = {2020},
  doi = {10.1103/PhysRevD.102.015027},
  eprint = {2006.00696},
  archivePrefix = {arXiv},
  primaryClass = {hep-th}
}

@article{BornInfeld1934,
  author = {Born, M. and Infeld, L.},
  title = {Foundations of the new field theory},
  journal = {Proc. Roy. Soc. Lond. A},
  volume = {144},
  pages = {425--451},
  year = {1934},
  doi = {10.1098/rspa.1934.0059}
}

@article{HeisenbergEuler1936,
  author = {Heisenberg, W. and Euler, H.},
  title = {Folgerungen aus der Diracschen Theorie des Positrons},
  journal = {Z. Phys.},
  volume = {98},
  pages = {714--732},
  year = {1936},
  doi = {10.1007/BF01343663}
}

@article{Schwinger1951,
  author = {Schwinger, J.},
  title = {On gauge invariance and vacuum polarization},
  journal = {Phys. Rev.},
  volume = {82},
  pages = {664--679},
  year = {1951},
  doi = {10.1103/PhysRev.82.664}
}

@article{Boillat1966,
  author = {Boillat, G.},
  title = {Vitesses des ondes \`electrodynamiques et lagrangiens exceptionnels},
  journal = {Ann. Inst. H. Poincare A},
  volume = {5},
  pages = {217--225},
  year = {1966}
}

@article{Boillat1970,
  author = {Boillat, G.},
  title = {Nonlinear electrodynamics: Lagrangians and equations of motion},
  journal = {J. Math. Phys.},
  volume = {11},
  pages = {941--951},
  year = {1970},
  doi = {10.1063/1.1665231}
}

@article{Novello2000,
  author = {Novello, M. and De Lorenci, V. A. and Salim, J. M. and Klippert, R.},
  title = {Geometrical aspects of light propagation in nonlinear electrodynamics},
  journal = {Phys. Rev. D},
  volume = {61},
  pages = {045001},
  year = {2000},
  doi = {10.1103/PhysRevD.61.045001},
  eprint = {gr-qc/9911085},
  archivePrefix = {arXiv}
}

@inproceedings{Novello2003,
  author = {Novello, M. and Perez Bergliaffa, S. E.},
  booktitle = {AIP Conference Proceedings},
  volume = {668},
  pages = {288--300},
  year = {2003},
  doi = {10.1063/1.1587103},
  eprint = {gr-qc/0302052},
  archivePrefix = {arXiv},
  primaryClass = {gr-qc}
}

@article{DeMelo2014,
  author = {de Melo, C. A. M. and Medeiros, L. G. and Pompeia, P. J.},
  title = {Causal structure and birefringence in nonlinear electrodynamics},
  journal = {Mod. Phys. Lett. A},
  volume = {30},
  number = {06},
  pages = {1550025},
  year = {2015},
  doi = {10.1142/S021773231550025X},
  eprint = {1407.0567},
  archivePrefix = {arXiv},
  primaryClass = {hep-th}
}

@article{RussoTownsend2023,
  author = {Russo, Jorge G. and Townsend, Paul K.},
  title = {Nonlinear electrodynamics without birefringence},
  journal = {JHEP},
  volume = {2023},
  number = {1},
  pages = {039},
  year = {2023},
  doi = {10.1007/JHEP01(2023)039},
  eprint = {2211.10689},
  archivePrefix = {arXiv},
  primaryClass = {hep-th}
}

@article{EscobarPotting2020,
  author = {Escobar, C. A. and Potting, R.},
  title = {Nonlinear vacuum electrodynamics and spontaneous breaking of {Lorentz} symmetry},
  journal = {Int. J. Mod. Phys. A},
  volume = {35},
  number = {27},
  pages = {2050174},
  year = {2020},
  doi = {10.1142/S0217751X20501742},
  eprint = {1810.01677},
  archivePrefix = {arXiv},
  primaryClass = {hep-th}
}

@article{Escobar2026MagneticCasimir,
  author = {Escobar, C. A. and Linares, Rom\'an and Mart\'in-Ruiz, A.},
  title = {Casimir effect near spontaneously Lorentz-breaking magnetic vacua in Pleba\'nski nonlinear electrodynamics},
  journal = {arXiv e-prints},
  eprint = {2606.00361},
  archivePrefix = {arXiv},
  primaryClass = {hep-th},
  year = {2026}
}

\end{document}